\documentclass[a4paper,11pt]{article}

\usepackage{contribution}

\newcommand{\weblink}[2][]{%
    \ifthenelse{\equal{#1}{}}%
    {\textnormal{\url{#2}}}%
    {\textnormal{\href{#2}{#1}}}%
}

\newcommand{\acknowledgements}[1]{%
  \bigskip\bigskip
  \textsf{\textbf{\Large Acknowledgements}} \\[2ex]
  {#1}
  \bigskip
}

\def\beq{\begin{equation}}
\def\eeq#1{\label{#1}\end{equation}}
\def\eeqn{\end{equation}}

\def\beqa{\begin{eqnarray}}
\def\eeqa#1{\label{#1}\end{eqnarray}}
\def\eeqan{\end{eqnarray}}

\let\bar=\overbar

\def\etal{{\it et al.}}

\def\Dslash{\not{\hbox{\kern-4pt $D$}}}
\def\dslash{\not{\hbox{\kern-2pt $\del$}}}

\def\msb{{\bar{\ssstyle M \kern -1pt S}}}

\newcommand{\contribution}[7][]{%
  \clearpage
  \thispagestyle{plain}
  \ifthenelse{\equal{#1}{}}
  {\hypersetup{pdftitle={#2}}}
  {\hypersetup{pdftitle={#1}}}
  \hypersetup{pdfauthor={{#3} {#4}}}
  {\centering\normalfont\LARGE\bfseries\sffamily #2 \par\nobreak}
  \lhead{}
  \chead{%
    \textit{\footnotesize XIV International Conference on Hadron Spectroscopy
      (\weblink[\textit{hadron2011}]{http://www.hadron2011.de}), 13-17 June 2011, Munich, Germany}%
  }
  \rhead{}
  \bigskip
  \begin{center}
    {#3} {#4}\ifthenelse{\equal{#6}{}}{}{\footnote{\weblink[#6]{mailto:#6}}}
    \ifthenelse{\equal{#7}{}}{}{#7} \\
    \textit{#5}
  \end{center}
  \bigskip
}

\renewcommand{\abstract}[1]{%
  \begin{center}
    \begin{minipage}{0.85\textwidth}
      \begin{footnotesize}
        #1
      \end{footnotesize}
    \end{minipage}
  \end{center}
  \bigskip
}

\begin{document}

%
%
%
%
%
%
{ 
\def\p{\pi}
\def\dkpp {$D^+ \! \to \! K^- \p^+ \p^+ $}
\def\dppp{$D^+  \to \pi^+\pi^+ \p^-$}
\def\ni {\noindent}
\def\nn {\nonumber}
\def\Kb{\bar{K}}
%
\contribution[Three-body FSIs in $D^+ \to K^- \p^+ \p^+$] 
{Three-body final state interactions in $D^+ \to K^- \p^+ \p^+$} 
{Patr\'{\i}cia C.}{Magalh\~{a}es} 
{$^1$ {\small Instituto de F\'{\i}sica, Universidade de S\~{a}o Paulo, S\~{a}o Paulo, SP, Brazil, 05315-970;} \\[-1.1mm] 
$^2$ {\small Instituto Tecnol\'ogico de Aeron\'autica, 
S\~ao Jos\'e dos Campos, SP, Brazil, 12.228-900;} \\[-1.1mm] 
$^3$ {\small Centro Brasileiro de Pesquisas F\'{\i}sicas, Rio de Janeiro, RJ, Brazil, 22290-180.}} 
{patricia@if.usp.br} 
{, M. R. Robilotta$^1$, K. S. F. F. Guimar\~{a}es$^2$, T. Frederico$^2$, W. de Paula$^2$, A. C. dos Reis$^3$, I. Bediaga$^3$. } 
%
\vspace*{-0.5cm}
\abstract{%
\vspace*{-0.5cm}
{\small We stress the importance of three-body final state interactions in $D^+ \to
K^- \p^+ \p^+$. The basic building block is the $K\pi$ amplitude with parameters determined by a fit to elastic LASS data. Based on a vector weak vertex, we can describe the $K\pi$ phase production experimental in the elastic region.
}
}
%

\section{Introduction}
Decays of $D$ mesons became an important source of information about  light scalars mesons, especially in the reactions \dkpp and \dppp. We calculate three-body effects in the decay \dkpp and our main motivation  is the discrepancy between the projection of $K^-\pi^+$ S-wave amplitudes from E791\cite{E791} and FOCUS\cite{FOCUS} experiments, and the scattering $K^-\pi^+$ S-wave from LASS\cite{LASS}.
To calculate the \dkpp decay, we need to deal with two independent families of processes: the weak vertex, usually treated by quark factorization techniques in the literature\cite{weak}, and the strong
final state interactions (FSIs), which do take place after the weak decay. 
We concentrate  on the three-body structure of
 FSIs and aim at identifying leading effects. Technical details of our calculation can be found in \cite{brazilian}. Among the simplifications made,
we mention the absence in the $K\pi$ amplitude of both isospin $3/2$ and $P$ waves,
as well as couplings to vector mesons and to inelastic channels. 

The $K\p$ amplitude is an essential ingredient in the three-body FSIs. We employ an elastic amplitude inspired on chiral perturbation
theory, eqs.(\ref{ea.2a}, \ref{ea.2}), supplemented by unitarization. The tuning to elastic LASS data\cite{LASS} defines the three free parameters in (\ref{ea.2a}, \ref{ea.2}).
This amplitude contains two poles, associated with the $\kappa$ and the $K^*_0(1430)$. 
\begin{eqnarray}
\bar{T}_{1/2} &\!=\! &
\frac{1}{F^2} \,[ s + 3\, t/4 - (M_\pi ^2 + M_K^2)]
- \;\frac{\alpha(s)}{s -  m_R^2} \;, 
\label{ea.2a}\\
\alpha &\!=\! & \frac{3}{2F^4} \;
[c_d\,s - (c_d - c_m) \, (M_\pi ^2 + M_K^2)\,]^2 \;.
\label{ea.2} 
\end{eqnarray}

In our exploratory work, we assume three
simple topologies for the weak amplitude,
indicated schematically in fig. \ref{FFSI-3}. The strengths of these vertices are respectively $W_a$,
$W_b$ and $W_c$,
taken as constants, and their strong evolution is studied independently.

\begin{figure}[htb]
\includegraphics[width=.99\columnwidth,angle=0]{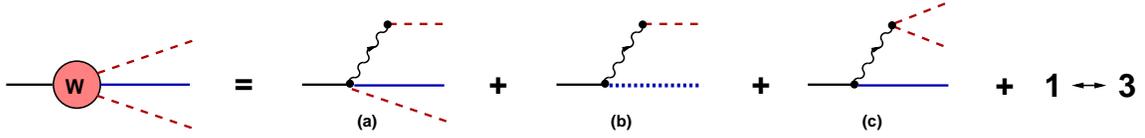}
\caption{ \small Topologies for the weak vertex: the dotted line is a scalar 
resonance and the wavy line is a $W^+$, which is contractile to a point 
in the calculation; in diagram $c$, one of the pions is neutral.}
\label{FFSI-3}
\end{figure}
\section{ Three-body FSIs}
Our treatment of FSIs departs from a Faddeev-like
integral equation, represented in fig.\ref{FPU-6} top, which is subsequently expanded perturbatively, fig.\ref{FPU-6} bottom.
\begin{figure}[htb]
\begin{center}
\includegraphics[width=0.8\columnwidth,angle=0]{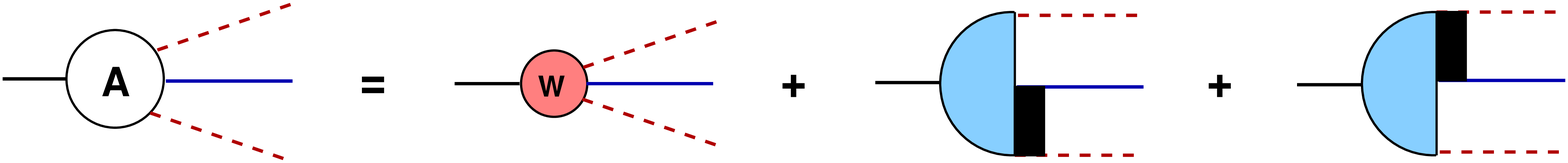}
\includegraphics[width=0.8\columnwidth,angle=0]{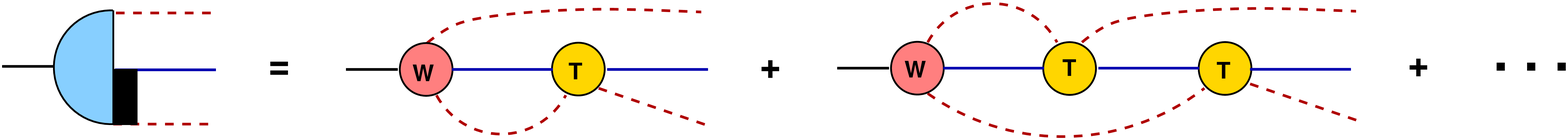}
\caption{ \small \dkpp decay: (top) partonic amplitude (red)
and hadronic multiple scattering in the ladder
approximation; (bottom) rescattering series implementing three-body unitary.}
\label{FPU-6}
\end{center}
\end{figure}
\ni Terms in the FSI series contain a recursive component, the $K\p$ two-body amplitude.
Each weak topology in fig.\ref{FFSI-3} is coupled to this series, giving rise to the amplitudes $A_a$ (fig. \ref{FPU-2}), $A_b$ (fig. \ref{FPU-3}) and $A_c$ (fig. \ref{FPU-4}).
Processes arising from the weak vertex $(a)$ in fig.\ref{FFSI-3} involves a tree term, whereas 
%
\begin{figure}[htb]
\begin{center}
\includegraphics[width=0.7\columnwidth,angle=0]{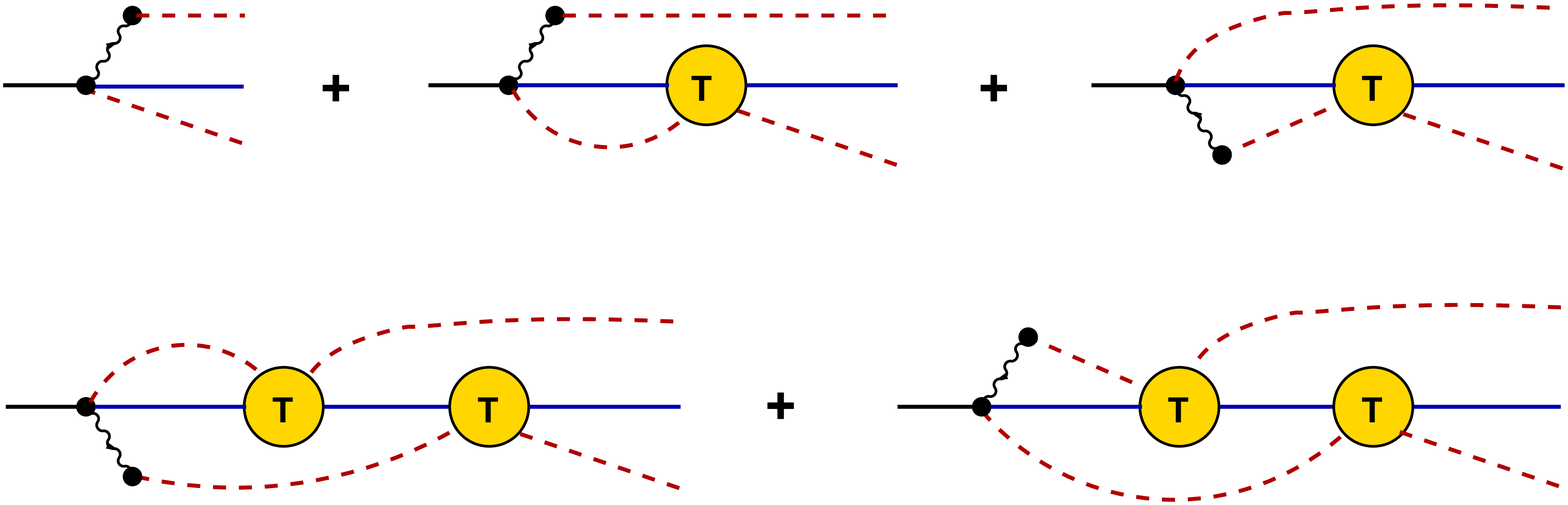}
\caption{ \small Diagrams involving the weak vertex $W_a$;
the wavy line is a $W^+$, always plugged to a $\p^+$;
the $\p$ produced together with the $\Kb$ on the opposite
side can be either positive or neutral.}
\label{FPU-2}
\end{center}
\end{figure}
the three-body rescattering series starting from the weak vertex
$b$ has to be treated properly in order to avoid double counting.
\begin{figure}[htb]
\begin{center}
\includegraphics[width=0.75\columnwidth,angle=0]{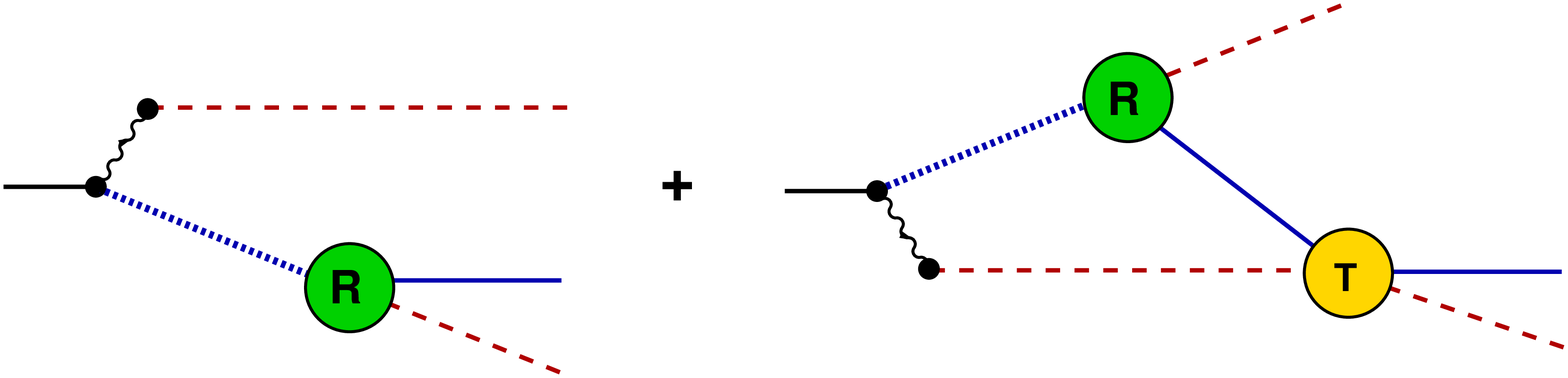}
\includegraphics[width=0.75\columnwidth,angle=0]{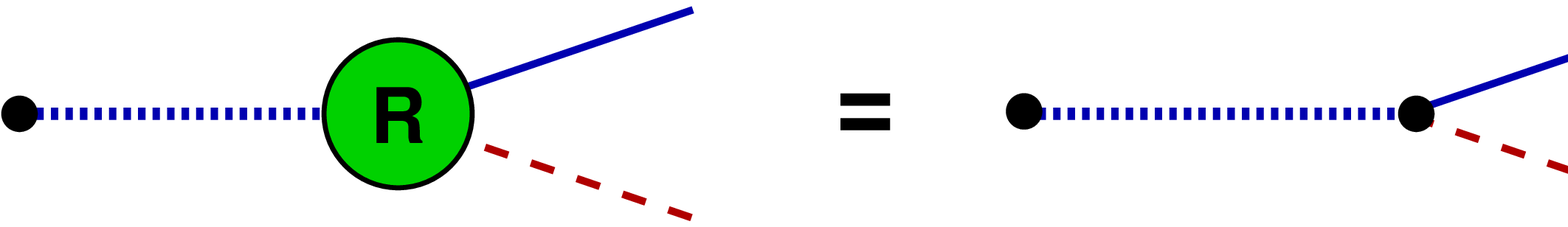}
\caption{ \small Diagrams involving the weak vertex $W_b$;
the wavy line is a $W^+$, always plugged to a $\pi ^+$
and the dotted line is a scalar resonance,
which has a width given by the substructure $R$
described at the bottom line.}
\label{FPU-3}
\end{center}
\end{figure}
\ni The bottom line in fig. \ref{FPU-3}  represents the construction of resonance width.
In the case of process associated with the weak vertex $c$ in fig.\ref{FPU-4}, the series
is simplified, since the $\pi_0$ produced directly from the $W^+$ decay is
not present in the final state and  the tree diagram does not play a role.
\begin{figure}[htb]
\begin{center}
\includegraphics[width=.85\columnwidth,angle=0]{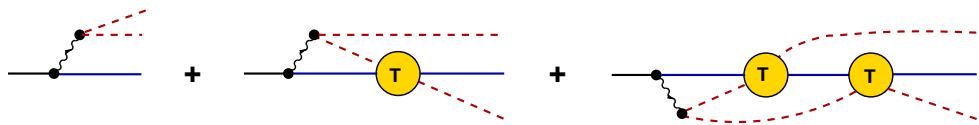}
\caption{ \small Diagrams involving $W_c$; one of the pions
in the weak vertex is neutral.}
\label{FPU-4}
\end{center}
\end{figure}

For simplicity,  we show here contributions from FSI series
 up to a single rescattering. This is a good approximation, since higher order terms tend to decrease. In the case of $A_a$, the leading contributions are given in the first line in fig.\ref{FPU-2}, which include both tree and one loop diagrams. In $A_b$, only the first diagram in the upper line of fig. \ref{FPU-3} is kept and, in $A_c$, the leading term is just the one loop diagram in the middle of fig. \ref{FPU-4}. In graph \ref{FPU-7}, we show these individual  contributions for the  phase, compared with the experimental scattering (LASS)\cite{LASS} and production (FOCUS)\cite{FOCUS} data. As we can see, while contributions from $A_a$ and $A_b$ fall exactly over the elastic $K\pi$ phase, the amplitude $A_c$ coincides with FOCUS data\cite{FOCUS} up to the region of the peak, when shifted by $-163^0$. The $A_c$ topology is the only involving only a proper three-body interactions whereas $A_a$ and $A_b$ include tree contributions.
\begin{figure}[htb]
\begin{center}
\includegraphics[width=0.85\columnwidth,angle=0]{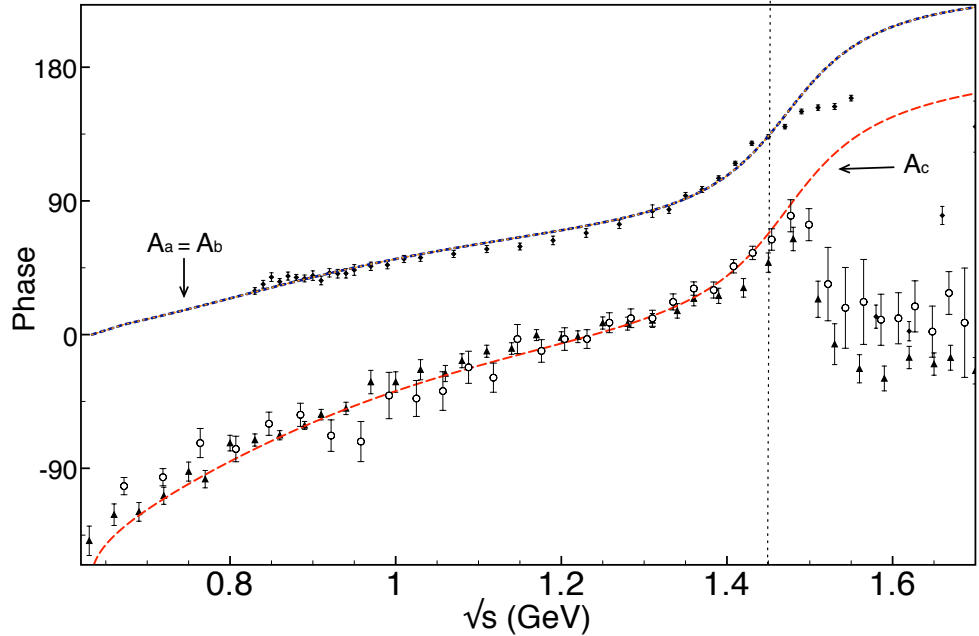}
\caption{ \small Leading contributions of the amplitudes $A_a$ (blue doted), $A_b$ (orange dashed) and $A_c$ shifted by $-163^0$ (red long dashed), compared with FOCUS \cite{FOCUS}(triangle), E791\cite{E791}(circle), and elastic $K\pi$ results from LASS\cite{LASS}(diamond).}
\label{FPU-7}
\end{center}
\end{figure}

Thus, with a simple model of three-body final state interactions and a number of simplifying assumption we can reconcile experimental data between two-body interactions and weak decays, stressing the importance of proper three-body effects in the \dkpp decay amplitude.
More calculation details can be found in reference\cite{brazilian}.


\acknowledgements{
This work was supported by FAPESP. PCM would like to thanks the organizers of Hadron2011 for the local support.
}


%
} 

\end{document}